\documentclass[10pt]{article}
\usepackage{graphicx,floatflt,amssymb,epsfig,epsf}
\def\issue(#1,#2,#3){{\bf #1}, #2 (#3)} 

\def\opcit(#1){ {\em op. cit.}, #1}
\def\etal {\em et al.}

\def\APP(#1,#2,#3){Acta Phys.\ Polon.\ \issue(#1,#2,#3)}
\def\ARNPS(#1,#2,#3){Ann.\ Rev.\ Nucl.\ Part.\ Sci.\ \issue(#1,#2,#3)}
\def\CPC(#1,#2,#3){Comp.\ Phys.\ Comm.\ \issue(#1,#2,#3)}
\def\CIP(#1,#2,#3){Comput.\ Phys.\ \issue(#1,#2,#3)}
\def\EPJC(#1,#2,#3){Eur.\ Phys.\ J.\ C\ \issue(#1,#2,#3)}
\def\EPJD(#1,#2,#3){Eur.\ Phys.\ J. Direct\ C\ \issue(#1,#2,#3)}
\def\IEEETNS(#1,#2,#3){IEEE Trans.\ Nucl.\ Sci.\ \issue(#1,#2,#3)}
\def\IJMP(#1,#2,#3){Int.\ J.\ Mod.\ Phys. \issue(#1,#2,#3)}
\def\JHEP(#1,#2,#3){J.\ High Energy Physics \issue(#1,#2,#3)}
\def\MPL(#1,#2,#3){Mod.\ Phys.\ Lett.\ \issue(#1,#2,#3)}
\def\NP(#1,#2,#3){Nucl.\ Phys.\ \issue(#1,#2,#3)}
\def\NIM(#1,#2,#3){Nucl.\ Instrum.\ Meth.\ \issue(#1,#2,#3)}
\def\PL(#1,#2,#3){Phys.\ Lett.\ \issue(#1,#2,#3)}
\def\PRD(#1,#2,#3){Phys.\ Rev.\ D \issue(#1,#2,#3)}
\def\PRL(#1,#2,#3){Phys.\ Rev.\ Lett.\ \issue(#1,#2,#3)}
\def\SJNP(#1,#2,#3){Sov.\ J. Nucl.\ Phys.\ \issue(#1,#2,#3)}
\def\ZPC(#1,#2,#3){Zeit.\ Phys.\ C \issue(#1,#2,#3)}
\def\JPG(#1,#2,#3){J.\ Phys.\ G \issue(#1,#2,#3)}

\def\bra {\langle}
\def\ket {\rangle}
\def\bfl{\begin{flushleft}}
\def\efl{\end{flushleft}}

\def\delms {\Delta {M_s}}
\def\l {\lambda}

\def\g {\gamma}
\def\r {\rightarrow}

\def\bar {\overline}
\def\bbbar {B^0-\bar{B^0}}

\def\bsbsbar {B_s-\bar{B_s}}
\def\msbar {\overline{MS}}

\def\be {\begin{equation}}
\def\ee {\end{equation}}
\def\bea {\begin{eqnarray}}
\def\eea {\end{eqnarray}}
\def\n {\nonumber}
\def\bc {\begin{center}}
\def\ec {\end{center}}

\begin{document}

\begin{flushright}
IMSC-PHYSICS/21-2006\\
CU-PHYSICS-11-2006\\
\end{flushright}

\begin{center}
{\Large \bf $\bsbsbar$ mixing, B decays and R-parity violating supersymmetry}\\
\vspace*{1cm}
\renewcommand{\thefootnote}{\fnsymbol{footnote}}
{\large {\sf Soumitra Nandi ${}^1$}
and {\sf Jyoti Prasad Saha ${}^2$}
} \\
\vspace{10pt}
{\small
${}^{1)}$ {\em Department of Physics, University of Calcutta, 92 A.P.C.
Road, Kolkata 700009, India}\\
${}^{2)}$ {\em Institute of Mathematical Sciences, Chennai 600113, India}}
\normalsize
\end{center}

\date{\today}

\begin{abstract}
We discuss the implications of the recent measurement of the $\bsbsbar$ 
oscillation frequency $\Delta M_s$ on the parameter space of R-parity violating
supersymmetry. For completeness, we also discuss the bounds coming from
leptonic, semileptonic, and nonleptonic B decay modes, and point out some
possibly interesting channels at LHC.
\end{abstract}

\bfl
\vspace*{0.05in}
\efl


\section{Introduction}

The $\bsbsbar$ mass difference, recently measured by the D0 \cite{D0} and 
the CDF \cite{CDF} collaborations, is given by,
\be
\label{Bsmix}
17~{\rm  ps}^{-1} <  \delms < 21~{\rm  ps}^{-1}\ \ 
  {(D0)},\ \ \ 
   \delms  = (17.31^{+0.33}_{-0.18} \pm 0.07)~{\rm ps}^{-1}
   {(CDF)}.
   \ee
This result is consistent with the Standard Model (SM) prediction, 
which is estimated as $21.3 \pm 2.6$ ps$^{-1}$ by the UTfit group \cite{utfit}
and as $20.9^{+4.5}_{-4.2}$ ps$^{-1}$ by the CKMfitter group 
\cite{ckmfitter}. The implications of $\delms$ measurements on the parameter
space of New Physics (NP)  have already been considered  
\cite{model_indep1,model_indep2,Buras:2006,Zprime}. 
However, given the hadronic uncertainties 
in the SM prediction, along with additional uncertainties when 
NP is included, the present measurement of 
$\delms$ does not provide a really strong constraint on NP \cite{model_indep1,
model_indep2}.

There have been some attempts to put bounds on the parameter space of R-parity
conserving supersymmetry (RPC SUSY) from the $\delms$ data \cite{susy}. 
In this paper, we would like to put bounds on the R-parity violating (RPV) SUSY 
couplings. We will use not only the $\delms$ data but also the data on the
leptonic, semileptonic, and nonleptonic branching ratios (BR) and CP
asymmetries of $B$ and $B_s$ mesons that are affected by the particular
RPV couplings. Such a work on $B^0$ mesons may be found in \cite{sk03} and
this is an extension of that work to the $B_s$ sector. For the relevant
formulae, we refer the reader to \cite{sk03}.

It has been shown in \cite{dreiner} that RPV couplings involving sleptons
($\l$ and $\l'$ type) generate nonzero neutrino mass and one can put
stringent constraints on them from the WMAP data \cite{wmap}. The exact
bounds depend on the relation of the mass matrices with the CKM matrix.
Anyway, such a study forces us to consider only those couplings which
can still be relatively large, and in this paper we derive better bounds 
on some of these product couplings than those coming from \cite{dreiner}.

A major motivation for this study is the $B_s$ physics that is going to be 
probed at LHC-b, and even at CMS or ATLAS during the low-luminosity run
of the Large Hadron Collider (LHC). The leptonic and semileptonic decays
are clean and any enhancement over the SM expectations will signify some
NP. Also, the phase $\chi$ in $\bsbsbar$ mixing comes in the subleading order
of the CKM matrix and is expected to be very small ($\sim 0.03$), 
so any CP-violating
effect in $B_s$ decays not involving an up quark will be interesting.

Effects of RPV SUSY on B physics have been discussed extensively in
the literature \cite{rpv:k,bsb,agashe,abel,rpv:b}. 
Constraints coming from neutral meson mixing have been discussed 
in \cite{sk03,gg-arc,sk02}. However, in these papers, the $B_s$ sector
could not be dealt with, since only the lower bound on $\delms$ existed then.
This paper, in a sense, is the completion of the series. All the 
computational details that have been taken into account in \cite{sk03,sk02}
({\em e.g.}, the NLO QCD corrections for short-distance effects, inclusion
of both SM and RPV) are incorporated in this paper. 

The paper is arranged as follows. In Section 2 we outline the relevant formulae
necessary for the analysis, and give the numerical inputs in Section 3.
The analysis on $\bsbsbar$ box and the decay processes is in Section 4, and
we conclude and 
summarize in Section 5. 

\section{Basic inputs}
\subsection{$\bsbsbar$ mixing}

The off-diagonal element $M_{12}$ in the $2\times 2$ effective Hamiltonian 
causes the $\bsbsbar$ mixing.  The mass difference
between the two mass eigenstates $\delms$ is given by (following
the convention of \cite{buras-fleischer})
\begin{equation}
\delms = 2|M_{12}|,
\end{equation}
with the approximation $|M_{12}|\gg |\Gamma_{12}|$ (this seems a good 
approximation even for the $B_s$ system).
If we have $n$ number of NP amplitudes with weak phases $\theta_n$, one can
write the mass difference between mass eigenstates as
\begin{equation}
\delms  =  2[ |M_{12}^{SM}|^2 + \sum_i|M_{12}^i|^2
 + 2|M_{12}^{SM}|\sum_i |M_{12}^i|\cos 2(\theta_{SM}-\theta_i)
 +  2 \sum_i \sum_{j>i} |M_{12}^j||M_{12}^i|\cos 2(\theta_j-\theta_i)
]^{1/2}.
\end{equation}

For $\bsbsbar$ system, the short-distance SM amplitude is
\be
M_{12}^{SM}\equiv {\bra \bar{B_s}|H_{eff}|B_s \ket\over 
2m_B}
= {G_F^2\over 6 \pi^2}(V_{ts}V_{tb}^*)^2
\eta_{B_s} m_{B_s} f_{B_s}^2 B_{B_s} m_W^2 S_0(x_t).
    \label{b-sm}
\ee
where generically $x_j = m_j^2/m_W^2$, $f_{B_s}$ is the $B_s$
meson decay constant, and $\eta_{B_s}$ and $B_{B_s}$
parametrize the short- and the long-distance QCD
corrections, respectively. 
The function $S_0$ is given by
\begin{equation}
S_0(x) = {4x-11x^2+x^3\over 4(1-x)^2} - {3x^3\ln x \over 2(1-x)^3}.
\end{equation}
If the NP amplitude has a nonzero phase, 
then there will be an effective phase in 
$\bsbsbar$ mixing amplitude, whose presence may be tested in the hadronic 
B factories.
In the presence of NP, the general $\Delta F = 2$ effective Hamiltonian 
can be written as 
\begin{equation}
{\cal H}_{eff}^{\Delta F = 2} = \sum_{i=1}^5 c_i(\mu) O_i(\mu) + \sum_{i=1}^3
\tilde c_i(\mu) \tilde O_i(\mu) + H.c.
\end{equation}
where $\mu$ is the regularization scale. The effective operators
$O_i$ and  $\tilde O_i$ are given in \cite{sk03}. 
The Wilson coefficients $c_i$ at $q^2=m_W^2$ include NP effects, coming from
couplings and internal propagators. However, for most of the NP models,
and certainly for the case we are discussing here, all NP particles are
heavier than $m_W$ and hence the running of the coefficients between $m_W$
and $\mu={\cal O}(m_b)$ are controlled by the SM Hamiltonian alone.
In other words, NP determines only the boundary conditions of the 
renormalization group (RG) equations.  
For the evolution of these coefficients down to the low-energy scale, we
follow Ref.\ \cite{becirevic}, which uses, for $\bbbar$ mixing, $\mu = m_b 
= 4.6$ GeV. 
The expectation values of these operators between $\bar{B_s}$ and $B_s$
at the scale $\mu$ are analogous to those as given in \cite{sk03}.
The $B_{B_s}(\mu)$ parameters have been taken from \cite{lattice-b}.

\subsection{R-parity violation}

It is well known that in order to avoid rapid proton decay one cannot
have both  lepton number and  baryon number violating RPV couplings, and we
shall work with a lepton number violating model. This leads
to both slepton (charged and neutral) and squark  
mediated decays, and new amplitudes for $\bsbsbar$ mixing, where particles 
flowing inside the box can be (i) charged sleptons and up-type
quarks, (ii) sneutrino and down type quarks, (iii) squarks and leptons. One
or both of the scalar particles inside the box can be replaced by 
$W$ bosons, charged Higgs bosons and Goldstone bosons (in a non-unitary gauge) 
(see Fig.\ 1). We follow the usual practice of avoiding the so-called ``pure
SUSY'' contributions to the box amplitudes, {\em i.e.}, those coming from
charginos, neutralinos or gluinos inside the loop. Not only the strongly
interacting superparticles are expected to be heavier than the electroweak
ones (and hence the contribution being suppressed), but also one can choose
SUSY models where these contributions become negligible ({\em e.g.}, alignment
in the squark sector, or Higgsino-dominated
lighter chargino, to kill off the respective boxes.) 
Since the current lower bound on the slepton mass is generally
weaker than that on squark mass by a factor 2-3, the slepton mediated
boxes have greater chance to be numerically significant.

We start with the superpotential
\begin{equation}
\label{w}
{\cal W}_{\lambda'} = \lambda'_{ijk} L_i Q_j D^c_k,
\end{equation}
where $i,j,k = 1,2,3$ are quark and lepton generation indices;
$L$ and $Q$ are the $SU(2)$-doublet lepton and quark superfields and
$D^c$ is the $SU(2)$-singlet down-type quark
superfield respectively. Written in terms of component fields, this
superpotential generates six terms, plus their hermitian conjugates:
\begin{equation}
\label{LQD}
{\cal L}_{LQD} = \l'_{ijk} \left[ \tilde\nu^i_L {\bar d}^k_R d^j_L
+ \tilde d^j_L {\bar d}^k_R \nu^i_L + (\tilde d^k_R)^* {\bar\nu}^i_L
d^j_L
- \tilde e^i_L {\bar d}^k_R u^j_L - \tilde u^j_L {\bar d}^k_R e^i_L
-(\tilde d^k_R)^* {\bar e }^i_L u^j_L\right] + H.c.
\end{equation}
With such a term, one can have two different kind of boxes, shown in Fig.\ 1,
that contribute
to $\bsbsbar$ mixing: first, the one where one has two sfermions flowing inside
the loop, alongwith two SM fermions \cite{decarlos-white}, and secondly,
the one where one slepton, one $W$ (or charged Higgs or Goldstone) and two
up-type quarks complete the loop \cite{gg-arc}. It is obvious that the first
amplitude is proportional to the product of four $\lambda'$ type
couplings, and the second to the product of two $\lambda'$ type couplings
times $G_F$. We call them L4 and L2 boxes, respectively, for brevity,
where L is a shorthand for $\l'$.

We will constrain only products of two $\lambda'$-type couplings at a time,
and assume a hierarchical structure, {\em i.e.}, only one product is,
for all practical purpose, simultaneously nonzero (but can have a nontrivial 
phase). This may not be physically 
the most appealing scenario but keeps the discussion free from unnecessary
complications. This product can in general be complex.
Any product is bounded by the product of the bounds on the individual terms,
which we call the direct product bound (DPB). Interesting
bounds are those which are numerically smaller, and hence stronger,
than the corresponding DPBs. The DPBs are mostly taken from \cite{dreiner},
and we highlight those products which are more tightly constrained than their
respective DPBs. The detailed formulae of the box amplitudes may be found
in \cite{sk03}. 

\begin{figure*}
\begin{center}
\centerline{\hspace*{3em}
\epsfxsize=14cm\epsfysize=4.5cm
                     \epsfbox{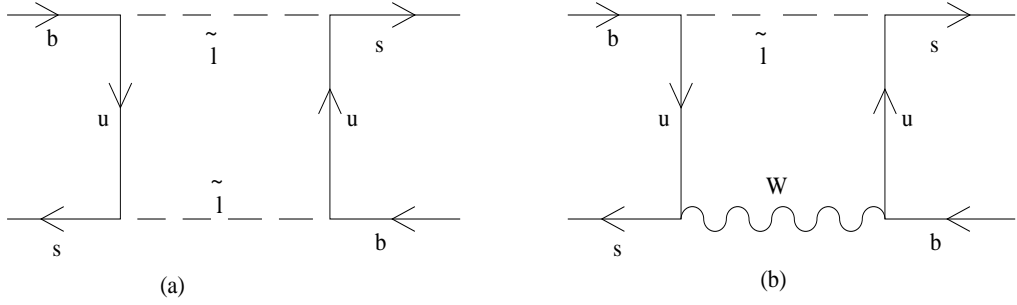}
}
\end{center}
\hspace*{-3cm}
  \caption{\em R-parity violating contributions to $\bsbsbar$ mixing.
Figure (a) corresponds to L4, while figure (b) to L2 amplitudes (see text for
their meanings). For L4, there are similar diagrams with squarks and leptons
(both charged and neutral), as well as diagrams with left-chiral quarks
as external legs and quarks and sneutrinos flowing in the box.
For L2, there are diagrams where
the $W$ is replaced by the charged Higgs or the charged Goldstone. The internal
slepton can be of any generation, and so can be the internal charge $+2/3$
quarks, generically depicted as $u$.}
\end{figure*}

\subsection{Semileptonic and leptonic decay channels}
The RPV couplings that may contribute to $\bsbsbar$ mixing should also
affect various $B$ decay modes. Let us first consider the leptonic and
semileptonic modes. 

The expected BRs of leptonic flavour conserving $\Delta B=1$ processes 
within SM are much below the experimental numbers (except $B\rightarrow 
K^{(*)}\ell^+\ell^-$), so one can safely ignore the SM effects as well as 
the R-conserving SUSY effects to put bounds on the RPV couplings. (The final
state leptons must be the same if the product coupling contributes to 
$\bsbsbar$ mixing.) The leptonic decay modes are theoretically clean and 
free from any hadronic uncertainties. The semileptonic modes have the usual
form-factor uncertainties.

To construct four-fermion operators from $\lambda'$ type couplings that mediate
such semileptonic and leptonic $B$ decays, one needs to integrate out the 
squark or slepton field. The product RPV coupling may in general be complex.
However, since all the leptonic decays are one-amplitude processes (only RPV,
for all practical purposes)
there is no scope for CP-violation; one can only look at nonzero BRs.
By the same argument, we can take all couplings to be real without any loss 
of generality. 


The effective Hamiltonian is of the form \cite{jyoti1}
\begin{eqnarray}
{\cal H}_{RPV} &=& {1\over 2} B_{jklm} \left[ \bar{\ell_j}\gamma^\mu P_L \ell_l
\right] \left[ \bar{u_m}\gamma_\mu P_R u_k\right] 
+ {1\over 2} B_{jklm} \left[ \bar{\nu_j}\gamma^\mu P_L \nu_l
\right] \left[ \bar{d_m}\gamma_\mu P_R d_k\right] \nonumber\\
&{}& - {1\over 2} C_{jklm} \left[ \bar{\nu_j}\gamma^\mu P_L \nu_l
\right] \left[ \bar{d_k}\gamma_\mu P_L d_m\right] + H.c.,
   \label{4fer}
\end{eqnarray}
where
\begin{equation}
B_{jklm} = \sum_{i=1}^3 \frac{{\l'}^*_{jik}\l'_{lim}}{m_{\tilde 
{L_i}}^2 },
C_{jklm} = \sum_{i=1}^3 \frac{{\l'}^*_{jki}\l'_{lmi}}{m_{\tilde d_{R_i}}^2 }.
\end{equation}
We take any one to be nonzero at a time. $m_{\tilde L_i}$ is the left-chiral
up/down squark mass, taken to be degenerate.

The entire leptonic $B_s$ decay amplitude is solely due to new physics, as far
as detectability is concerned. In RPV models, squark-mediated $\l'\l'$ type 
interactions are responsible for such purely leptonic decays. As already 
pointed out, the bounds are robust in a sense that they are free from any 
theoretical uncertainties (except for the decay constants of $B_s$), and do 
not depend on the phase of the RPV couplings.
For the $B$ mesons, no such leptonic mode has yet been observed. The 
corresponding upper limits on the BRs are of the order of
$10^{-7}$ for $\ell=\mu$ and $10^{-5}$ for $\ell=e$ modes. With 300 GeV
squarks, from the bounds that one obtains here, a BR at most
of the order of $10^{-8}$ can be expected. Thus, we do not expect to see such 
leptonic channels before the next-generation hadronic machine or super $e^+e^-$
$B$ factories. However, a number of semileptonic modes $b\to s\ell^+\ell^-$
have been observed and the BRs are at the SM ballpark.

The decay width of $B_{s}\r \ell^-\ell^+$ is given in \cite{jyoti1}. For the
semileptonic decays, we use
\bea
\bra K(p_2) | \bar{s}\gamma^\mu b|B(p_1)\ket &=&
P^\mu F_1(q^2) +q^{\mu} \frac{m_B^2-m_K^2}{q^2}
\left(F_0(q^2)-F_1(q^2)\right),
\nonumber\\
\langle \phi(p_2,\epsilon ) |\ V_\mu \mp A_\mu \ |B_s(p_1)
         \rangle &=&\frac{1}{m_{B_s}+m_{\phi}}
         [-iV(q^2) \varepsilon _{\mu \nu \alpha \beta }
         \epsilon ^{*\nu }P^{\alpha }q^{\beta }\nonumber \\ 
	 &&\pm A_0 (q^{2})(P\cdot q)
          \epsilon _{\mu }^{*} \pm A_+ (q^2) (\epsilon ^* \cdot p_1)P_{\mu }
           \nonumber \\
        && \pm A_- (q^2) (\epsilon ^*\cdot p_1)q_{\mu }] 
\label{Bphi} 
\eea
 where $m_{B_s}$ and $m_{\phi}$ are the meson masses,
 $p_1 (p_2)$ is the momentum of the initial (final) meson,
 $\epsilon$ is the polarization vector of the vector meson 
 $\phi$, $P=p_1+p_2$, $q=p_1-p_2$, $V_\mu= \bar{q}_2\gamma _{\mu } q_1$, 
 $A_\mu=\bar{q}_2\gamma _{\mu } \gamma_5 q_1$, $V$, $A_{0,\pm}$ and 
 $ F_{(1/0)}(q^2)$ are the form factors. 
 The values of these form factor are taken from \cite{gengliu,ali1}.

The RPV matrix elements for the decay mode  $B_{(s/d)}\to (\phi/K) l^+ l^-$ 
(where $l= e,\mu$):
\bea
{\cal M}(B_{(s/d)} \r (\phi/K) \ell_i^+ \ell_i^-)&=&{1\over 2} B_{i3i2} 
\left( \bar{\ell^+}\gamma^\mu P_L \ell^- \right)
\left( \langle (\phi/K)| \bar{(s/d)}\gamma_\mu P_R b|B_{(s/d)}\rangle \right) 
\eea

\subsection{ Nonleptonic decay channels:}

There are four types of slepton-mediated nonleptonic decays that can proceed
through the relevant RPV couplings. Among them, $b\to c\bar{c}s$ and $b\to
s\bar{s}s$ rates are bound to be undetectably small if these couplings are
to be compatible with the neutrino mass bounds \cite{dreiner}. We, therefore,
focus on the $b\to u\bar{u}s$ and $b\to d\bar{d}s$ type transitions. They
mediate the channels $B\to\pi K$, $B\to\rho K$. The corresponding $B_s$ decay
channels do not have data at comparable level. Note that both BRs and CP
asymmetries for these channels have been measured \cite{babar,belle}. 
While the data does not uniquely point to NP, the trend is encouraging.

We will use the Conventional Factorization (CF) model \cite{ali1} to analyze
the effect of RPV SUSY on the $\pi K$ and $\rho K$ channels. While the 
validity of such a simple approach may be questioned, it is not wildly off
the truth, at least for these channels. The effective Hamiltonian reads
\bea
H_{RPV} &=&      d^R_{jkn} (\bar d_n \g^\mu P_L d_j)_8
                                   (\bar d_k \g_\mu P_R b)_8
           +  d^L_{jkn} (\bar d_n \g^\mu P_L b)_8
                                   (\bar d_k \g_\mu P_R d_j)_8\n\\
          &{ }& +  u^R_{jnk} (\bar u_n \g^\mu P_L u_j)_8
                                   (\bar d_k \g_\mu P_R b)_8
\eea
where
\be
d^R_{jkn} = \sum_i {\l'_{ijk}{\l'}^*_{in3} \over 2m_{\tilde\nu_{Li}}^2 },
\ \ d^L_{jkn} = \sum_i {\l'_{i3k}{\l'}^*_{inj} \over 2m_{\tilde\nu_{Li}}^2 },
\ \ u^R_{jnk} = \sum_i {\l'_{ijk}{\l'}^*_{in3} \over 2m_{\tilde  e_{Li}}^2 }.
   \label{nonlep1}
\ee

Following the standard practice we shall assume that the RPV 
couplings are hierarchical {\em i.e.}, only one combination of the 
couplings is numerically significant.
Let us assume, to start with, 
only $d^L_{112}$ and $u^R_{112}$ to be
nonzero. The QCD corrections are easy to implement:
the short-distance QCD corrections
enhance the $(S-P)\times(S+P)$ RPV operator by approximately
a factor of 2 while running from the slepton mass scale 
(assumed to be at 100 GeV) to $m_b$ \cite{bagger}. 

The RPV amplitude for $B\r\pi K$ is given by
\be
M_{\pi^0 K^-} = {1\over {\sqrt{2}}}[u^R_{112} ({-R_1(A_{\pi K}^{(1)})
+A_{\pi K}^{(2)}{1\over N_c}}) + d^L_{112}{1\over N_c}(A_{\pi K}^{(2)})] 
\ee
\be
M_{\pi^0 \bar{K^0}} = {1\over {\sqrt{2}}}[d^L_{112}( {-R_1(A_{\pi K}^{(1)})
+A_{\pi K}^{(2)}{1\over N_c}}) + u^R_{112}{1\over N_c}(A_{\pi K}^{(2)})]
\ee
\be
M_{\pi^- \bar{K^0}} ={d^L_{112}}({R_1} A_{\pi K}^{(1)})
\ee
\be
M_{\pi^+ K^-} ={u^R_{112}}({-R_1} A_{\pi K}^{(1)})
\ee
The expressions for the $B\r\rho K$ amplitudes will be similar to those shown
above, with $R_1$ replaced by $R_2$ and $A^{(i)}_{\pi K}$ replaced by
$A^{(i)}_{\rho K}$. We use the shorthand
\be
R_1 = {2}{m_{K^0}^2\over (m_u+m_s) (m_b-m_u)},\ \ 
R_2 = {2}{m_{\rho}^2\over (m_u+m_s) (m_b-m_u)},
\ee
and
\bea
&& A_{\pi K}^{(1)} = f_{K} F_0^{B\r\pi}(m_K^2) (m_B^2-m_{\pi}^2),\ \ 
A_{\pi K}^{(2)} = f_{\pi} F_0^{B\r K}(m_\pi^2) (m_B^2-m_K^2),
\nonumber\\
&&A_{\rho K}^{(1)} = 2 f_{K}  m_{\rho} A_0^{B\r\rho}(m_K^2)
(\epsilon_{\rho}.p_K),\ \
A_{\rho K}^{(2)} = 2 f_{\rho}  m_{\rho} F_1^{B\r K}(m_\rho^2)
(\epsilon_{\rho}.p_K).
\eea

\section{Numerical Inputs}

\begin{table}[htbp]
\begin{center}
\begin{tabular}{||c|c||}
\hline
Quantity & Value \\
\hline
$\delms $ & $(17.31^{+0.33}_{-0.18} \pm 0.07) ps^{-1}$ \\
$\gamma$ & $50^\circ$-$72^\circ$ \\
$\eta_{B_s}$   & $0.55\pm 0.01$ \\
$m_t^{\bar{MS}}(m_t^{\bar{MS}})$ & 166 GeV \\
$m_b^{\bar{MS}}(m_b^{\bar{MS}})$ & 4.23 GeV \\
$m_b(m_b)$                       &  4.6 GeV \\
$m_c(m_b)$                       &  1.3 GeV \\
$m_d(m_b)$                       &  5.4 MeV \\
$m_s(2~{\rm GeV})$               &  125 MeV \\
$f_B\sqrt{B_{B_s}}|_{JLQCD}$ & $(0.245\pm 0.021^{+0.003}_{-0.002}) $ GeV \\
$|V_{us}|\times 10^1$ & $2.272_{-0.01}^{+0.01}$\\
$|V_{cs}|\times 10^1$ & $9.73$ \\
$|V_{ts}|\times 10^3$ & ${41.61}_{-0.78}^{+0.12} $ \\
$|V_{ub}|\times 10^3$ & $4.4 \pm 0.3$ \\
$|V_{cb}|\times 10^3$ & $42.0\pm 0.7$ \\
$|V_{tb}|\times 10^1$ & $9.99$ \\
\hline
\end{tabular}
\caption{Input parameters used for the numerical analysis, from \cite{CDF,
ckmfitter,becirevic,JLQCD}. }
\label{inputtable}
\end{center}
\end{table}

The major sources of the numerical inputs are: (i) the Heavy Flavor
Averaging Group (HFAG) website \cite{hfag} for the latest (summer 2006)
updates on $B$ physics; 
(ii) Particle Data Group 2006 edition \cite{pdg2006}; and
(iii) the inputs used in the CKMfitter package \cite{ckmfitter}. 
The quark masses and Wilson coefficients have been taken from \cite{becirevic,
ciuchini}. We use the following numbers.

The masses for all the mesons $B^0$, $B^-$, $\pi$, $\rho$, and $K$ are 
the corresponding central values as given in \cite{pdg2006}.
The meson decay constants (in GeV) are:
\be
f_\pi = 0.133,\ f_K = 0.158,\ f_\rho = 0.210.
\ee
The transition formfactors \cite{bsw} at $q^2=0$ are given by
\be
F_0(B\r K) = 0.38;\ F_0(B\r\pi) = 0.33;\ A_0(B\r\rho) = 0.28,
\ee
and $F_0(0) = F_1(0)$. 

The quark masses have been evaluated in the $\msbar$ scheme. The pole
mass for the top quark is about 5 GeV higher and the mass for the bottom
quark is $4.6$ GeV. The CKM elements are shown in Table 1.  


The leptonic and semileptonic BRs for the $B$ meson, which are
of interest to us, are as follows \cite{D0,pdg2006,hfag}:
\begin{eqnarray}
{\rm Br} (B\r K\ell^+\ell^-) &<& (0.57\pm 0.07)\times 10^{-6}\ \  
(\ell=e/\mu);\nonumber\\
{\rm Br} (B\r Ke^+e^-) &=&(0.55\pm 0.09)\times 10^{-7}; \nonumber\\
{\rm Br} (B\r K\mu^+\mu^-)&=& (0.61\pm 0.08)\times 10^{-7};\nonumber\\
{\rm Br}(B_s\to \phi \mu^+ \mu^-) & < & 4.1\times 10^{-6};
\nonumber\\
{\rm Br}(B_s \to \mu^+\mu^-) & < & 1 \times 10^{-7};
\nonumber\\
{\rm Br}(B_s \to  e^+ e^-) & < & 5.4\times 10^{-5};
\nonumber
\end{eqnarray}


For the $\pi K$ and $\rho K$modes, the data reads 
\cite{pdg2006,hfag,aleksan,babar0111087,belle0207033}:
\bea
Br(B^0\r\pi^- K^+) &=& (18.9\pm 0.7)\times 10^{-6}\nonumber\\
Br(B^0\r\pi^0 K^0) &=& (11.5\pm 1.0)\times 10^{-6}\nonumber\\
Br(B^+\r\pi^+ K^0) &=& (24.1\pm 1.3)\times 10^{-6}\nonumber\\
Br(B^+\r\pi^0 K^+) &=& (12.1\pm 0.8)\times 10^{-6}\nonumber\\
Br(B^0\r \rho^- K^+) &=& (9.9^{+1.6}_{-1.5})\times 10^{-6}\nonumber\\
Br(B^0\r\rho^0 K^0) &=& (5.1\pm {1.6})\times 10^{-6}\nonumber\\
Br(B^+\r\rho^0 K^+) &=& (4.23^{+0.56}_{-0.57})\times 10^{-6}\nonumber\\
A_{CP}^{dir}(B^0\r\pi^\pm K^\mp) &=& 0.115\pm 0.018.
\eea

To evaluate the QCD corrections, we take $\alpha_s(m_Z^2)$ = 0.1187
\cite{pdg2006}, and take the SUSY scale $M_S = 500$ GeV. The precise
value of this scale is not important, however, and we can take it to be at
the squark mass scale (300 GeV) without affecting the final results. The
exact evolution matrix can be found in \cite{becirevic} and \cite{ciuchini};
for our purpose, it is sufficient to note that for the $B_s$ system, 
the operator
$\tilde{O_1}$ is multiplicatively renormalized by a factor 0.820 at the 
scale $\mu=2$ GeV, and the operator $O_4$ at $m_W$ changes to 
$(2.83  O_4 + 0.08  O_5)$. 
We again stress that theoretically the procedure is questionable
for boxes with light quarks flowing in the loop. However, the numbers that
we obtain are fairly robust and one can very well drop the NLO
corrections altogether, if necessary, without compromising the results.
Since the $O_5$ admixture is small, one can take the central values
for these parameters without introducing too much error.
The relevant $B$-parameters for $B_s$ system are taken from \cite{lattice-b}

We take all sleptons to be degenerate at 100 GeV, and all
squarks at 300 GeV. We also take $\tan\beta (\equiv v_2/v_1) = 5$ (very low
values are excluded by LEP, and the numbers are not sensitive to the precise
choice of $\tan\beta$), and the charged Higgs boson mass as 200 GeV (lower 
values are disfavored from $b\to s\gamma$).

\section{Analysis}
\subsection{$\bsbsbar$ mixing}
For the $B_s$ system, the bounds are summarized in Table 2. When the product
coupling is complex, we show only the real part, since the bound on the imaginary
part is almost equal to this. The reason is easy to understand: the bounds 
are obtained when the RPV coupling has a phase opposite to that of the
SM coupling, so that the interference is destructive. At the limit where
the RPV coupling determines the mixing amplitude, the phase is irrelevant.
The effect of this destructive interference is clear in Fig. 2(a) and Fig. 2(b).
For a more detailed explanation, we refer the reader to \cite{sk03}. 

\begin{table}
\begin{center}
\begin{tabular}{||c|c|c||}
\hline
$\l'\l'$ & Only & Complex, \\
combination & real & real part \\
\hline
(i32)(i33)&$1.01 \times 10^{-2}$&$1.0 \times 10^{-2}$\\
\hline
(i22)(i23) & $8.2 \times 10^{-3}$ & $8.0 \times 10^{-3}$ \\
\hline
(i12)(i13) & $1.2 \times 10^{-2}$ & $3.7 \times 10^{-2}$ \\
\hline
(i22)(i33) & $1.8 \times 10^{-1}$ & $1.7 \times 10^{-1}$  \\
\hline
(i32)(i23) & $3.2 \times 10^{-4}$ & $3.0 \times 10^{-4}$  \\
\hline
(i22)(i13) & $3.47 \times 10^{-2}$ & $3.45 \times 10^{-2}$ \\
\hline
(i12)(i23) &  $5.16 \times 10^{-2}$ & $7.56 \times 10^{-2}$\\
\hline
(i32)(i13) & $1.4 \times 10^{-3}$ & $1.3 \times 10^{-3}$ \\
\hline
(i12)(i33) & $9.0 \times 10^{-1}$ & $9.0 \times 10^{-1}$ \\
\hline
(i23)(i33) & $1.66 \times 10^{-2}$ & $5.1 \times 10^{-2}$ \\
\hline
(i22)(i32) & $1.66 \times 10^{-2}$ & $5.1 \times 10^{-2}$ \\
\hline
(i21)(i31) & $1.66 \times 10^{-2}$ & $5.1 \times 10^{-2}$  \\
\hline
\hline
\end{tabular}
\caption{Bounds on $\l'\l'$ combinations from $\bsbsbar$ mixing.
The table displays the magnitudes only, and not the signs.
}
\end{center}
\end{table}

\begin{figure*}[htbp]
\vspace{-10pt}
\centerline{\hspace{-3.3mm}
\rotatebox{-90}{\epsfxsize=6cm\epsfbox{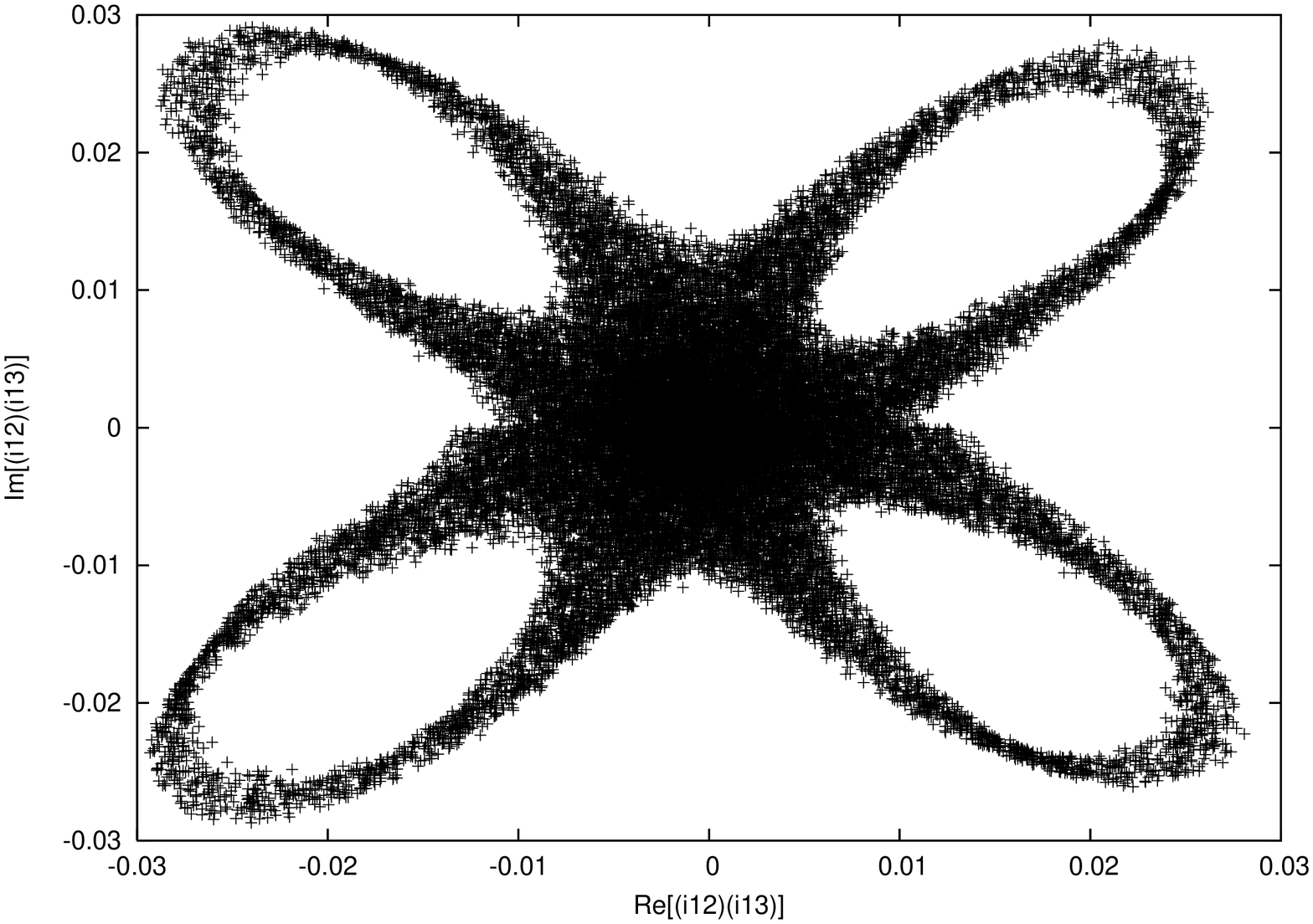}}
\hspace{-0.1cm}
\rotatebox{-90}{\epsfxsize=6cm\epsfbox{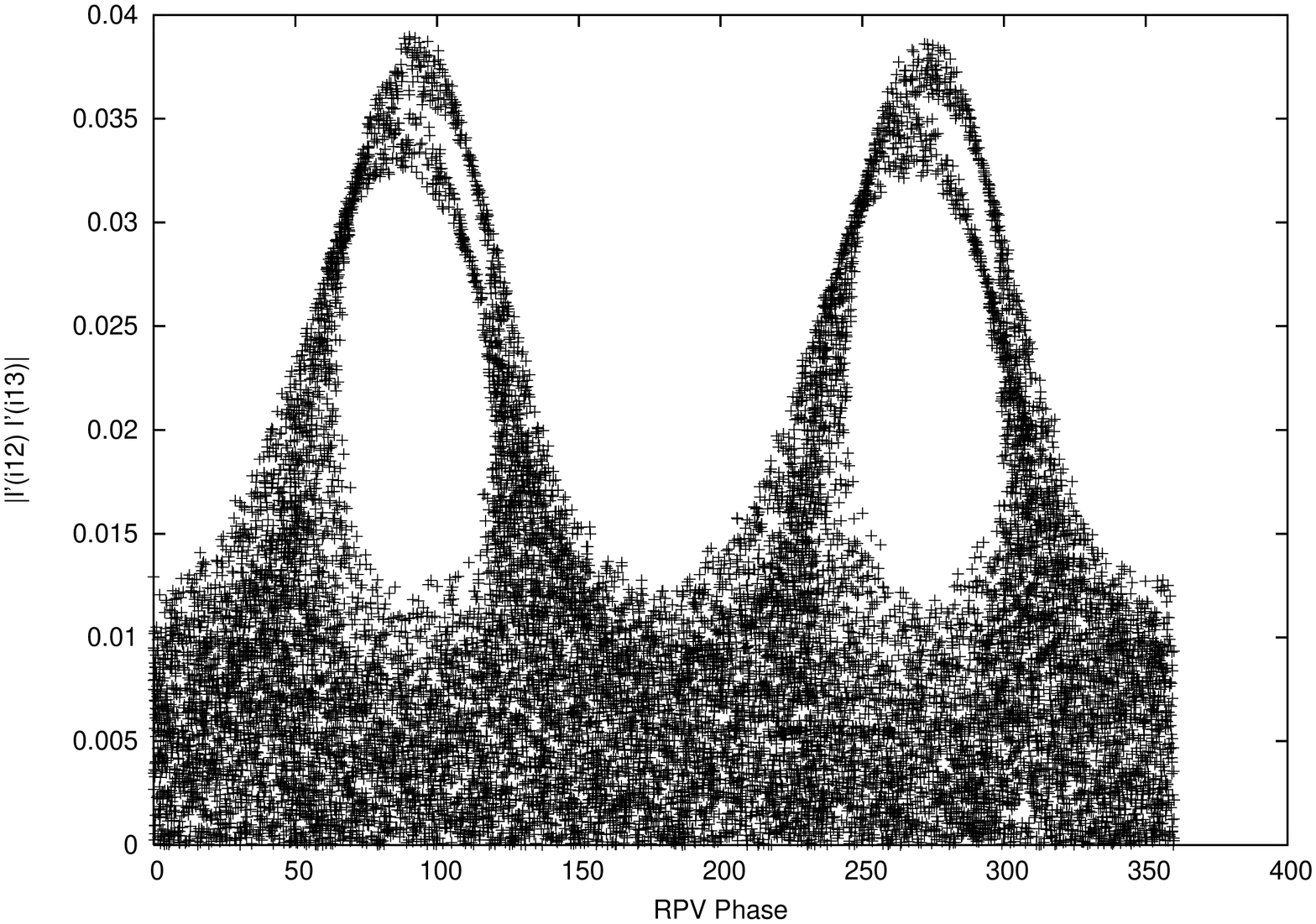}}}
\vspace*{3mm}
\centerline{\hspace{-0.5cm} (a) \hspace{7.5cm} (b)}
\hspace{3.3cm}
\caption{(a) Allowed parameter space for $\l'_{i12}\l'_{i13}$
(b) The allowed paramater space of RPV phase for 
$|\l'_{i12}\l'_{i13}|$,which can gives CP assymetries in the 
$B_s \r K \pi$ decay.}
\end{figure*}
The relative magnitudes of the bounds are also easy to understand.
For example let us consider the bounds on $\l'_{i32}\l'_{i23}$ vis-a-vis 
$\l'_{i22}\l'_{i33}$.
The relevant box diagrams have the same particle content; but the first one
is proportional to $V_{ts}V_{cb}$ ($\sim {\cal O}(\lambda^4)$), 
and the second one to $V_{tb}V_{cs}$ ($\sim $1). The relative suppression 
in $\lambda$ enhances the limit on the RPV coupling. 

Though most of the bounds are of the same order in magnitude, these are,
theoretically, an improvement over those obtained earlier \cite{gg-arc,
decarlos-white,sk02}.We have taken into account all possible amplitudes 
(and the interference
patterns play a nontrivial role), including the SM one, but have systematically
neglected the pure supersymmetric boxes coming from gaugino exchange. The
reason is that those boxes decouple in the heavy squark limit, and one can
always take an RPV model embedded in a minimal supersymmetric theory where 
such FCNC processes are somehow forbidden. It was shown in \cite{sk02} that
the bounds are fairly robust even if one takes into account such SUSY 
contributions. Furthermore, the QCD corrections are implemented upto NLO.  

\subsection{Nonleptonic decay channels}
\begin{table}
\begin{center}
\begin{tabular}{||c|c|c|c||}
\hline
$\l'\l'$ & Quark & Meson & Bound\\
combination & level & level & \\
\hline
(i21)(i31) & $b\to d\bar{d} s$ & $B_d \to \pi^0 \bar {K^0}$ 
&$2.45 \times 10^{-3}$\\
& &$B^- \to \pi^0  K^-$ &\\
\hline
(i12)(i13) & $b\to d\bar{d} s$ & $B^- \to \pi^0 K^-$ 
&$2.82 \times 10^{-3}$\\
& $b\to u\bar{u} s$ & $B_d \to \pi^+ K^-$ &\\
\hline
(i22)(i32) & $b\to s\bar{s} s$ & $B_s \to \phi  K_s$ 
&$2.33 \times 10^{-3}$\\
\hline
\end{tabular}
\caption{Some of the possible nonleptonic transitions mediated by
the RPV couplings discussed in the paper.}
\end{center}
\end{table}

As we have mentioned, the $B\to\pi K$ and $B\to\rho K$ numbers are 
encouraging for NP enthusiasts. However, the hadronic uncertainties are
significant. Also, one must have a nonzero strong phase between the SM tree 
and the SM penguin amplitudes to explain the direct CP asymmetry data on
$B\to\pi^+K^-$. Thus, it is of importance to explore the data in conjunction
with $\bsbsbar$ mixing. In our analysis, we use the CF model, as mentioned,
and vary the strong phase difference between the SM tree and the RPV
amplitudes from 0 to $2\pi$. To take into account the hadronic uncertainties,
we (i) vary the SM amplitude from its CF value by 20\%, and (ii) vary the
RPV weak phase in the range [0:$2\pi$]. The allowed parameter space of
RPV couplings, in the magnitude-phase plane, comes out with separate
island-like structures.
For example, 
$\l'_{i31}\l'_{i21}$ lies between $2.05 \times 10^{-3}$ to
$2.45 \times 10^{-3}$ and the corresponding phase lies between 
$104^\circ $ to $120^\circ$, whereas $\l'_{i13}\l'_{i12}$ lies between 
$2.31 \times 10^{-3}$ to $2.82 \times 10^{-3}$ and the corresponding 
phase lies between $85^\circ $ to $105^\circ$. We quoted the highest value in 
Table 3. These bounds are much stronger then that coming from 
$\bsbsbar$ mixing. 

We have not considered the $B\r(\eta,\eta') K$ modes. Though they are 
mediated by $b\r s\bar{q}q$ ($q=u,d$) transitions, the BRs 
for those modes cannot be explained simultaneously by such 
a simple new physics structure \cite{sn-js-ak}.
Similarly, the mode $B\r\phi K^*$ has not been considered, since
the longitudinal polarization anomaly for this mode cannot be explained 
without a 
contribution from tensor current but RPV SUSY do not provide for
such tensor current structures, at least at the tree-level \cite{sn-pd}. 
However, we note that unless a product coupling is at least of the order of
$10^{-3}$, the RPV contribution is unlikely to affect the SM amplitude.

\subsection{Leptonic and semileptonic decay channels}

The detail procedure of the leptonic and semileptonic decay channels are
given in \cite{jyoti1}. While their bounds were for 100 GeV squarks, we scale
the numbers for 300 GeV squarks. The bounds are shown in Table 4. Note that
the channel $B\to K\mu^+\mu^-$ gives the best bound. Also, there is no bound
involving $\tau$s in the final state, but we can estimate the number of 
$B_s\to\tau^+\tau^-$ decays. The relevant RPV coupling is $\l'_{3i2}\l'_{3i3}$.
It can easily be checked that unless $i=1$, the product is so constrained
from neutrino mass \cite{dreiner} that even at LHC-b, there is no hope to
detect a RPV signal in this channel. For $i=1$, the coupling $\l'_{312}
\l'_{313}$ should be less than $2.8\times 10^{-3}$, which in turn translates
into a bound on the BR to be less than $2.7\times 10^{-6}$. Note that the
SM expectation is about $7\times 10^{-7}$.

\begin{table}
\begin{center}
\begin{tabular}{||c|c|c|c||}
\hline
$\l'\l'$ & Quark & Meson & Bound\\
combination & level & level & \\
\hline
(2i2)(2i3) & $b\to s \mu^+ \mu^-$& $B_s \to \mu^+ \mu^-$ 
&$7.7 \times 10^{-3}$\\
\hline
(2i2)(2i3) & $b\to s \mu^+ \mu^-$& $B_s \to \phi \mu^+ \mu^-$ 
&$4.9 \times 10^{-3}$\\
\hline
(2i2)(2i3) & $b\to s \mu^+ \mu^-$& $B \to K \mu^+ \mu^-$ 
&$6.6 \times 10^{-4}$\\
\hline
(1i2)(1i3) & $b\to s e^+ e^-$& $B \to K e^+ e^-$ 
&$7.7 \times 10^{-4}$\\
\hline
\end{tabular}
\caption{Some of the possible leptonic and semileptonic transitions mediated 
by the RPV coupling relevent with $\bsbsbar$ mixing are given here. It has
shown that in many cases the bounds coming from this decays are much better 
then coming from mixing.}
\end{center}
\end{table}

\begin{table}
\begin{center}
\begin{tabular}{||c|c|c|c||}
\hline
$\l'\l'$ & Related  & Current  & Previous\\
combination& process &bound  & bound\\
\hline
(112)(113)&$B_s \to K e^+ e^- $&$7.74 \times 10^{-4}$&
${1.52\times 10^{-1}}_{\nu}$\\
\hline
(122)(123)&$B_s \to K e^+ e^- $&$7.74 \times 10^{-4}$&
${9.7\times 10^{-6}}_{\nu}$\\
\hline
(132)(133)&$B_s \to K e^+ e^- $&$7.74 \times 10^{-4}$&
${6.0\times 10^{-5}}_{\nu}$\\
\hline
(212)(213)&$B_s \to K \mu^+ \mu^- $&$6.57 \times 10^{-4}$&
${1.52\times 10^{-1}}_{\nu}$\\
\hline
(222)(223)&$B_s \to K \mu^+ \mu^- $&$7.74 \times 10^{-4}$&
${9.7\times 10^{-6}}_{\nu}$\\
\hline
(232)(233)&$B_s \to K \mu^+ \mu^- $&$7.74 \times 10^{-4}$&
${6.0\times 10^{-5}}_{\nu}$\\
\hline
(312)(313)&$B^- \to \pi^0 K^-$ & $2.8 \times 10^{-3}$ 
&${1.52\times 10^{-1}}_{\nu}$ \\
&$B_d \to \pi^+ K^-$&&\\ 
\hline
(322)(323)&$\bsbsbar $ & $8.2 \times 10^{-3}$ 
&${9.7\times 10^{-6}}_{\nu}$ \\
\hline
(332)(333)&$\bsbsbar $ & $1.01 \times 10^{-2}$ 
&${6.0\times 10^{-5}}_{\nu}$ \\
\hline
(i22)(i33) &$\bsbsbar$ &$1.8 \times 10^{-1}$ & ${3.75\times 10^{-9}}_{\nu}$ \\
\hline
(i32)(i23)$\dag$&$\bsbsbar$&$3.2 \times 10^{-4}$ & ${1.56\times 10^{-1}}_{\nu}$ \\
\hline
(i22)(i13) &$\bsbsbar$& $3.47 \times 10^{-2}$  & ${9.75\times 10^{-6}}_{\nu}$  \\
\hline
(i12)(i23) &$\bsbsbar$&  $5.16 \times 10^{-2}$ & ${1.52\times 10^{-1}}_{\nu}$ \\
\hline
(i32)(i13) &$\bsbsbar$& $1.4 \times 10^{-3}$ & ${1.56 \times 10^{-1}}_{\nu}$ \\
\hline
(i12)(i33) &$\bsbsbar$& $9.0 \times 10^{-1}$ & ${5.8 \times 10^{-5}}_{\nu}$  \\
\hline
(i23)(i33) &$\bsbsbar$& $1.66 \times 10^{-2}$ & ${5.8\times 10^{-5}}_{\nu}$ \\
\hline
(i22)(i32) &$B_s \to \phi K_s$& $2.3 \times 10^{-3}$ 
& ${1.0\times 10^{-5}}_{\nu}$ \\
\hline
(i21)(i31) &$B_d \to \pi^0 \bar {K}^0$& $2.45 \times 10^{-3}$ 
& ${1.56 \times 10^{-1}}_{\nu}$ \\
&$B^- \to \pi^0 \bar {K}^-$ & &\\
\hline
\hline
\end{tabular}
\caption{Bounds on real $\l'\l'$ combinations from $\bsbsbar$ mixing and
correlated decay channels. The DPBs, displayed in the last column, occur from
neutrino constraints with no mixing scenario \cite{dreiner}.
The product marked with a dagger is bounded from tree-level $\bsbsbar$
mixing($\sim {\cal O} (10^{-6})$).
}
\end{center}
\end{table}

\subsection{Comparison between bounds coming from mixing and decay}

In Table 5 we summarize our results, displaying the bounds on all $\l'\l'$
type products that may be responsible for $\bsbsbar$ mixing. We find that 
a number of them may have better bounds from semileptonic or nonleptonic decay
modes. While the neutrino constraints are indeed tight, we obtain a 
tighter constraint for most of the products.

\section{Summary and Conclusions}

In this paper we have computed the bounds on the product couplings of the 
type $\l'\l'$ coming from  $\bsbsbar$ mixing. Though such an analysis is not 
new, we have implemented several features in the analysis which have not 
been taken into account in earlier studies. Previously there was a lower 
limit on $\delms$, here we have used the current bound on it and considered 
the exact expression for the box amplitude taking all possible processes, 
including that from SM. The QCD corrections to the amplitudes have been 
taken upto the NLO level. We have considered the possibility that the RPV 
product couplings may be complex. The analysis is done in the benchmark point 
$m_{H^+}=200$ GeV, $\tan\beta = 5$, all sleptons degenerate at 100 GeV and 
all squarks degenerate at 300 GeV, and neglecting the pure MSSM contribution 
to the box amplitudes (by possibly applying some underlying FCNC suppression 
principle, like alignment of the squark mass matrices). 

It is to be observed that in some cases, our bounds are actually weaker than 
those obtained earlier by saturating the mass difference with RPV alone. 
The reason is that destructive interference with the SM amplitude plays a very 
crucial role in determining the bounds, particularly when the phase of the RPV 
coupling is arbitrary. There is an intricate interplay among different 
amplitudes as can be seen in Fig. 2. 

In some cases the bounds obtained from semileptonic $B$ decays are better. 
Of course, one can enhance the squark mass to a limit where these bounds 
become weaker then those obtained from the box (the latter is not much 
affected by decoupling the squarks), but such extremely massive squarks are 
not interesting, even for the LHC. 

However, some of these couplings may affect the nonleptonic decay modes 
(which, being slepton mediated, cannot be suppressed by decoupling the 
squarks). For some cases, the bounds coming from such decays are tighter
than those coming from mixing.

\section{Acknowledgements}
We would like to thank Rahul Sinha and Anirban Kundu for their useful comments
and suggestions,

\end{document}